\documentclass[a4paper]{article}
\usepackage[left=20mm,right=20mm,top=20mm,bottom=25mm]{geometry}
\usepackage[affil-it]{authblk}
\usepackage{amsmath}
\usepackage{amssymb}
\usepackage{tikz}
\usepackage{hyperref}
\usepackage{microtype}
\usepackage{multirow}
\usepackage{cite}

\usetikzlibrary{arrows}
\usetikzlibrary{decorations.pathmorphing}
\usetikzlibrary{decorations.markings}
\usetikzlibrary{calc}

\tikzset{
  >=stealth',
  midarrow/.style={
    postaction={
      decorate,
      decoration={markings, mark=at position .55 with {\arrow{>}}}
    }
  },
  fermion/.style=midarrow,
  photon/.style={
    decorate,
    decoration={snake, amplitude=2pt, segment length=8pt}
  },
  boson/.style={
    decorate,
    decoration={snake, amplitude=2pt, segment length=8pt}
  },
  gluon/.style={
    decorate,
    decoration={coil, amplitude=4pt, segment length=5pt}
  },
  scalar/.style=densely dashed,
  arrowsnake/.style={
    preaction={photon, draw},
    postaction=midarrow
  }
}

\hypersetup{
  colorlinks,
  linkcolor={blue!50!black},
  citecolor={blue!50!black},
  urlcolor={blue!50!black}
}

\newcommand{\abs}[1]{\lvert #1 \rvert}
\newcommand{\nuc}[2]{$^{#2}${#1}}
\newcommand{\pvec}[1]{\vec{#1}\mkern2mu\vphantom{#1}}

\begin{document}

\title{Equivalent photons in proton-proton and ion-ion collisions at~the~LHC}

\author[1,2,3]{M.~I.~Vysotsky \thanks{vysotsky@itep.ru}}
\author[1,3]{E.~V.~Zhemchugov \thanks{zhemchugov@itep.ru}}

\affil[1]{\small Institute for Theoretical and Experimental Physics, 117218,
Moscow, Russia}
\affil[2]{\small National Research University Higher School of Economics,
101978, Moscow, Russia}
\affil[3]{\small Moscow Engineering Physics Institute, 115409, Moscow, Russia}

\date{}

\maketitle

\begin{abstract}
  Equivalent photon approximation is used to calculate fiducial cross sections
  for dimuon production in ultraperipheral proton-proton and lead-lead
  collisions. Analytical formulae taking into account experimental cuts are
  derived. The results are compared with the measurements reported by the ATLAS
  collaboration.
\end{abstract}

\section{Introduction}

This year we celebrate the 111th anniversary of L.~D.~Landau. This paper is
devoted to the modern state of the problem first considered by L.~D.~Landau and
E.~M.~Lifshitz in~1934 when they calculated the production cross section of $e^+
e^-$ pair in ultrarelativistic heavy ions collisions~\cite{pzs6.244}. We will
demonstrate that this problem is still of great interest.

In spite of many efforts, no New Physics has been found at the LHC so far. It
might be a good time to consider scenarios of appearance of New Physics at the
LHC that were less attractive at the time when the LHC was under construction.
Although the LHC was conceived as a hadron-hadron collider, it also acts as a
photon-photon collider with the photons appearing in ultraperipheral collisions
of hadrons. This idea is quite old, and it was thoroughly considered during the
construction and operation of the RHIC and the LHC~\cite{pr163.299,
hep-ph-0112211, hep-ph-0112239, hep-ex-0201034, hep-ph-0304301, nucl-ex-0502005,
hep-ph-0611042, 0706.3356, 0810.1400, 1104.0571, 1404.0896, 1601.07001,
1607.05095, 1610.06647, 1708.09836}.
However, since hadronic interactions were more likely to deliver the signal of
New Physics, particularly in Higgs boson properties, they received more
attention in the literature and were given higher priority in the LHC schedule.
With the long shutdown of the LHC beginning at the end of 2018, it might be a
good time to reconsider photon-photon collisions at the LHC as a source of
possible New Physics events so that the necessary detectors adjustments could
be made and, perhaps, more time for heavy ions collisions could be negotiated in
the LHC schedule. 

\begin{figure}[t]
  \centering
  \begin{tikzpicture}
    \coordinate (P1in)  at (-1.7,  0.7);
    \coordinate (V1)    at (-0.7,  0.7);
    \coordinate (P1out) at ( 1.0,  1.7);
    \coordinate (V)     at (   0,    0);
    \coordinate (P2in)  at (-1.7, -0.7);
    \coordinate (V2)    at (-0.7, -0.7);
    \coordinate (P2out) at ( 1.0, -1.7);

    \draw [fermion] (P1in) node [left] {$\text{Pb}$} -- (V1);
    \draw [fermion] (V1)   -- (P1out) node [right] {$\text{Pb}$};
    \draw [fermion] (P2in) node [left] {$\text{Pb}$} -- (V2);
    \draw [fermion] (V2)   -- (P2out) node [right] {$\text{Pb}$};
    \draw [photon]  (V1)   -- node [above right] {$\gamma$} (V);
    \draw [photon]  (V2)   -- node [below right] {$\gamma$} (V);

    \draw (V) -- +( 15:1);
    \draw (V) -- +(  5:1);
    \draw (V) -- +( -5:1);
    \draw (V) -- +(-15:1);

    \draw [fill=gray] (V) circle (0.2);
  \end{tikzpicture}
  \caption{Feynman diagram for an ultraperipheral lead-lead collision.}
  \label{f:pb-pb}
\end{figure}
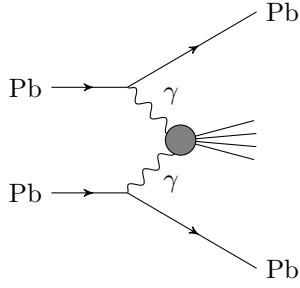

The leading order Feynman diagram for an ultraperipheral collision is presented
in Fig.~\ref{f:pb-pb} where instead of lead nuclei there could be any charged
particles. The distinctive signature of an ultraperipheral collision is that the
charged particles remain intact after the collision. These particles won't have
high transverse momentum, so they are difficult to detect with just the main
detectors of the ATLAS and CMS experiments, but there exist additional detectors
at low scattering angles (the ATLAS forward proton
detector~\cite{cern-lhcc-2015-009} and the CMS-TOTEM precision proton
spectrometer~\cite{cern-lhcc-2014-021}). However, even without the forward
detectors, ultraperipheral collisions manifest through production of particles.

Let us compare proton-proton and lead-lead ultraperipheral collisions as
possible sources of New Physics events. Integrated luminosity delivered by the
LHC in Run~2 in proton-proton collisions is over 150~fb$^{-1}$ both for the
ATLAS and the CMS experiments.  Integrated luminosity delivered by the LHC in
lead-lead collisions in the heavy ions run was
$0.7~\text{nb}^{-1}$ in 2015~\cite{tumpw027} and $1.8~\text{nb}^{-1}$ in
2018~\cite{cms-luminosity}. Cross section for an ultraperipheral
collision is proportional to $Z^4$ where $Z$ is the particle charge. For Pb, $Z
= 82$, so we get that if there exists New Physics that appears in $\gamma
\gamma$ collisions, there will be $(150~\text{fb}^{-1}) / (82^4 \cdot
2.5~\text{nb}^{-1}) \approx 1.3$ times more events of it during the whole Run~2
$pp$ collisions than there were during the whole heavy ions run. However,
Run~2 duration was over 500~days (not counting the 2015 when only
$4.2~\text{fb}^{-1}$ were delivered in $pp$ collisions), while the heavy ions
run has lasted about 20 days in 2015 and 25 days in 2018. The $Z^4$ enhancement
of the cross section makes the search of New Physics in ultraperipheral
collisions of heavy ions at the LHC to look very promising.

The common approach to calculate cross sections of particles production in
ultraperipheral collisions is to use the equivalent photon
approximation (EPA)~\cite{zeit.phys.29.315, zeit.phys.88.612, williams,
pzs6.244} (see also~\cite{jetp11-388, rev.mod.phys.4.615, pepan4.239,
prep15.181}).  To compare the result with the experimental data, fiducial cross
section has to be calculated, which is the total cross section after applying
the experimental cuts on the phase space designed to reduce the background and
to take into account detector blind spots.  The fiducial cross section is
usually calculated from the total cross section with the help of the Monte Carlo
method (see, e.g., the {\tt SuperCHIC} MC generator~\cite{1508.02718}). The
equivalent photon approximation makes it possible to apply the most common
experimental cuts analytically, so often no Monte Carlo method is required.

In this paper, the equivalent photon approximation is used to calculate the
cross section of the $pp (\gamma \gamma) \to pp \mu^+ \mu^-$ reaction.
Then three kinds of experimental cuts are applied in succession:
\begin{enumerate}
  \item The cut on the invariant mass of muon pair $\sqrt{s}$: $\hat
  s_\text{min} < s < \hat s_\text{max}$.
  \item The cut on muon transverse momentum $p_T$: $p_T > \hat p_T$.
  \item The cut on muon pseudorapidity $\eta$: $\abs{\eta} < \hat \eta$.
\end{enumerate}
Numerical values of these cuts vary from experiment to experiment and from
measurement to measurement. The result of the calculation is used to obtain the
theoretical description for the experimental values provided by the ATLAS
collaboration~\cite{1708.04053}.  In this measurement, $\hat s_\text{min}$ was
chosen to be 12~GeV to avoid contributions from vector meson decays into $\mu^+
\mu^-$ (the heaviest of vector mesons belong to the $\Upsilon$ family); $\hat
s_\text{max} = 70$~GeV; $\hat p_T$ is 6 or 10~GeV depending on the invariant
mass; $\hat \eta$ is 2.4 so that the muon will hit the muon spectrometer.

The same formulae are used to calculate the fiducial cross section for the
reaction $\text{Pb} \; \text{Pb} \; (\gamma \gamma) \to \text{Pb} \; \text{Pb}
\; \mu^+ \mu^-$ studied in Ref.~\cite{atlas-conf-2016-025}. 

\section{Cross section of the $\mu^+ \mu^-$ production without cuts}

\label{s:xsection}

The distribution of equivalent photons generated by a moving particle with the
charge $Z e$ is
\begin{equation}
  n(\vec q) \mathrm{d}^3 q
  = \frac{Z^2 \alpha}{\pi^2}
    \frac{\vec q_{\perp}^{\, 2}}{\omega q^4}
    \mathrm{d}^3 q
  = \frac{Z^2 \alpha}{\pi^2 \omega}
    \frac{\vec q_{\perp}^{\, 2}}
         {(\vec q_{\perp}^{\, 2} + (\omega / \gamma)^2)^2}
    \mathrm{d}^3 q,
  \label{epa-distribution}
\end{equation}
where $q$ is the photon 4-momentum, $\vec q_\perp$ is its transverse component,
$\omega$ is the photon energy, $\gamma$ is the Lorentz factor of the particle.
For a proton with the energy $E = 6.5$~TeV, $\gamma = E / m_p \approx 6.93 \cdot
10^3$. To obtain the equivalent photon spectrum, one has to integrate this
expression over the transverse momentum up to some value~$\hat q$. The value of
$\hat q$ should be chosen so that the parent particle does not break apart when
emitting a photon of such momentum. For the proton, $\hat q = 0.20$~GeV (see
Appendix~\ref{s:form-factor} for derivation). Hence, the equivalent photon
spectrum is
\begin{equation}
  n(\omega) \mathrm{d} \omega
  = \frac{2 Z^2 \alpha}{\pi}
    \ln \left( \frac{\hat q \gamma}{\omega} \right)
    \frac{\mathrm{d} \omega}{\omega}
  \label{spectrum}
\end{equation}
in the limit $\omega \ll \hat q \gamma$. This simple expression allows us to
obtain analytical formulas for the cross section of muon pair production with
the experimental cuts.

\begin{figure}[!tbh]
  \centering
  \begin{tikzpicture}

    \coordinate (A)   at (-0.7,  1.2);
    \coordinate (B)   at (   0,  0.5);
    \coordinate (C)   at (   0, -0.5);
    \coordinate (D)   at (-0.7, -1.2);
    \coordinate (P1i) at (-1.2,  1.2);
    \coordinate (P1o) at (   1,  1.2);
    \coordinate (P2i) at (-1.2, -1.2);
    \coordinate (P2o) at (   1, -1.2);
    \coordinate (M1)  at (   1,  0.5);
    \coordinate (M2)  at (   1, -0.5);

    \draw[fermion] (P1i) node [left] {$p$} -- (A);
    \draw[fermion] (A)   -- (P1o) node [right] {$p$};
    \draw[fermion] (P2i) node [left] {$p$} -- (D);
    \draw[fermion] (D)   -- (P2o) node [right] {$p$};
    \draw[photon]  (A)   -- (B);
    \draw[photon]  (D)   -- (C);
    \draw[fermion] (M1)  node [right] {$\mu$} -- (B);
    \draw[fermion] (B)   -- (C);
    \draw[fermion] (C)   -- (M2) node [right] {$\mu$};

    \draw [->] ([yshift=-2mm]$(A)!0.1!(B)$)
               to node [midway, below left] {$(\omega_1, \vec q_1)$}
               ([yshift=-2mm]$(A)!0.6!(B)$);
    \draw [->] ([yshift=2mm]$(D)!0.2!(C)$)
               to node [midway, above left] {$(\omega_2, \vec q_2)$}
               ([yshift=2mm]$(D)!0.7!(C)$);
  \end{tikzpicture}
  ~
  \begin{tikzpicture}
    \coordinate (A)   at (-0.7,  1.2);
    \coordinate (B)   at (   0,  0.5);
    \coordinate (C)   at (   0, -0.5);
    \coordinate (D)   at (-0.7, -1.2);
    \coordinate (P1i) at (-1.2,  1.2);
    \coordinate (P1o) at (   1,  1.2);
    \coordinate (P2i) at (-1.2, -1.2);
    \coordinate (P2o) at (   1, -1.2);
    \coordinate (M1)  at (   1,  0.5);
    \coordinate (M2)  at (   1, -0.5);

    \draw[fermion] (P1i) node [left] {$p$} -- (A);
    \draw[fermion] (A)   -- (P1o) node [right] {$p$};
    \draw[fermion] (P2i) node [left] {$p$} -- (D);
    \draw[fermion] (D)   -- (P2o) node [right] {$p$};
    \draw[photon]  (A)   -- (C);
    \draw[photon]  (D)   -- (B);
    \draw[fermion] (M1)  node [right] {$\mu$} -- (B);
    \draw[fermion] (B)   -- (C);
    \draw[fermion] (C)   -- (M2) node [right] {$\mu$};

    \draw [->] ([xshift=-2mm]$(A)!0.2!(C)$)
               to node [midway, left] {$(\omega_1, \vec q_1)$}
               ([xshift=-2mm]$(A)!0.5!(C)$);
    \draw [->] ([xshift=-2mm]$(D)!0.2!(B)$)
               to node [midway, left] {$(\omega_2, \vec q_2)$}
               ([xshift=-2mm]$(D)!0.5!(B)$);
    
  \end{tikzpicture}  
  \caption{Leading order Feynman diagrams for the $pp (\gamma \gamma) \to pp
  \mu^+ \mu^-$ reaction.}
  \label{f:pp-diagram}
\end{figure}
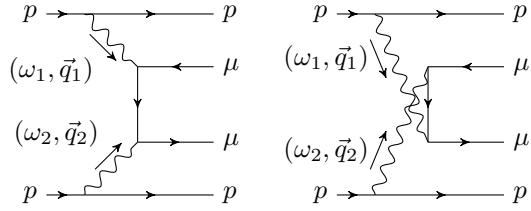

\begin{figure}[!tb]
  \centering
  \begin{tikzpicture}
    \draw [->, very thin] (-0.4, 0) -- (2.7, 0) node [below] {$\omega_1$};
    \draw [->, very thin] (0, -0.4) -- (0, 2.7) node [left]  {$\omega_2$};
    \draw (2, 0) node [below] {$\hat q \gamma$}
       -- (2, 2)
       -- (0, 2) node [left] {$\hat q \gamma$};

    \draw [dashed] plot [domain=0.11:2.7, samples=100]
          (\x, {0.3/\x});
    \draw [thick] plot [domain=0.37:2.7, samples=100]
          (\x, {1/\x}) node [right] {$s = 4 \omega_1 \omega_2$};
    \draw [thick] (0, 0)
               -- (2.7, 1.8) node [right] {$x = \frac{\omega_1}{\omega_2}$};
  \end{tikzpicture}
  \caption{Integration domain of~\eqref{pp-nocuts-omega}. The dashed line
  corresponds to $s = 4 m_\mu^2$. The domain is above the dashed line and inside
  the square.}
  \label{f:integration-space}
\end{figure}
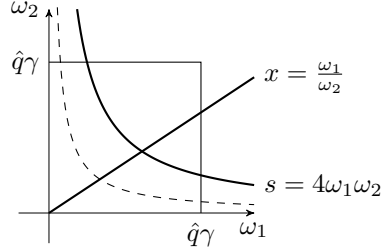

Muon pair production in ultraperipheral proton-proton collisions in the leading
order is described by the Feynman diagrams in Fig.~\ref{f:pp-diagram}.
The corresponding cross section is
\begin{equation}
  \sigma(pp (\gamma \gamma) \to pp \mu^+ \mu^-)
  = \int\limits_{m_\mu^2 / \hat q \gamma}^{\hat q \gamma}
      \mathrm{d} \omega_1
      \int\limits_{m_\mu^2 / \omega_1}^{\hat q \gamma}
        \mathrm{d} \omega_2 \,
        \sigma(\gamma \gamma \to \mu^+ \mu^-) \,
        n(\omega_1) \, n(\omega_2),
  \label{pp-nocuts-omega}
\end{equation}
where $\omega_1$ and $\omega_2$ are the photon energies and $\sigma(\gamma
\gamma \to \mu^+ \mu^-)$ is the Breit-Wheeler cross section~\cite{pr46.1087}:
\begin{equation}
  \sigma(\gamma \gamma \to \mu^+ \mu^-)
  = \frac{4 \pi \alpha^2}{s}
    \left[
      \left( 1 + \frac{4 m_\mu^2}{s} - \frac{8 m_\mu^4}{s^2} \right)
      \ln \frac{1 + \sqrt{1 - 4 m_\mu^2 / s}}
               {1 - \sqrt{1 - 4 m_\mu^2 / s}}
      - \left( 1 + \frac{4 m_\mu^2}{s} \right)
        \sqrt{1 - \frac{4 m_\mu^2}{s}}
    \right],
\end{equation}
$s = 4 \omega_1 \omega_2$ is the invariant mass of the muons. The integration
domain of~\eqref{pp-nocuts-omega} is presented in
Fig.~\ref{f:integration-space}.  It is convenient to change the integration
variables from $\omega_1$, $\omega_2$ to $s$ and $x$ where $x = \omega_1 /
\omega_2$.
Then the integration can be rearranged as follows:
\begin{equation}
  \begin{split}
    \sigma(pp (\gamma \gamma) \to pp \mu^+ \mu^-)
    &= \int\limits_{(2 m_\mu)^2}^{(2 \hat q \gamma)^2}
          \mathrm{d} s \,
          \sigma(\gamma \gamma \to \mu^+ \mu^-)
          \int\limits_{s / (2 \hat q \gamma)^2}^{(2 \hat q \gamma)^2 / s}
             \frac{\mathrm{d} x}{8 x} \,
             n \left( \sqrt{\frac{sx}{4}} \right)
             n \left( \sqrt{\frac{s}{4x}} \right)
    \\
    &= \frac{\alpha^2}{2 \pi^2}
       \int\limits_{(2 m_\mu)^2}^{(2 \hat q \gamma)^2}
          \frac{\mathrm{d} s}{s} \,
          \sigma(\gamma \gamma \to \mu^+ \mu^-)
          \int\limits_{s / (2 \hat q \gamma)^2}^{(2 \hat q \gamma)^2 / s}
             \frac{\mathrm{d} x}{x}
             \ln \frac{(2 \hat q \gamma)^2}{s x}
             \ln \left[ \frac{(2 \hat q \gamma)^2}{s} \cdot x \right]
  \end{split}
  \label{pp-nocuts}
\end{equation}
(note the symmetry of the integral under the $x \to 1 / x$ replacement). Thus,
we get
\begin{equation}
  \sigma(pp (\gamma \gamma) \to pp \mu^+ \mu^-)
  = \frac{16 \alpha^2}{3 \pi^2}
    \int\limits_{(2 m_\mu)^2}^{(2 \hat q \gamma)^2}
      \frac{\mathrm{d} s}{s} \, 
      \sigma(\gamma \gamma \to \mu^+ \mu^-)
      \ln^3 \frac{2 \hat q \gamma}{\sqrt{s}}.
  \label{pp-nocuts-ll}
\end{equation}

Since $\sigma(\gamma \gamma \to \mu^+ \mu^-)$ falls as $1 / s$ for $s \gg 4
m_\mu^2$, in the leading logarithmic approximation the logarithm in this
expression should be taken at $s = 4 m_\mu^2$. Then\footnote{
  An incorrect spectrum of equivalent photons was used
  in~\cite[Eq.~(1.4)]{prd4.1532} ($\ln^2(E/m_e)$ should be replaced with
  $\ln(E/\omega_1) \ln(E/\omega_2)$ inside the integral), which resulted into an
  extra factor of $3/2$ in Eq.~(5.4) and note [23] in the same paper. This
  error was later propagated into~\cite[Eq.(5.4)]{rev.mod.phys.4.615}.
  See~\cite[the second footnote on page~256]{prep15.181} for the discussion of
  similar errors often occurring in the applications of the equivalent photon
  approximation.}
\begin{equation}
  \sigma(pp (\gamma \gamma) \to pp \mu^+ \mu^-)
  = 8 \cdot \frac{28}{27}
    \frac{\alpha^4}{\pi m_\mu^2}
    \ln^3 \frac{\hat q \gamma}{m_\mu}.
  \label{pp-nocuts-landau}
\end{equation}
In this formula, when the masses of the produced particles $m$ are considerably
less than $\hat q$, the latter should be replaced with $m$.\footnote{In the case
of $\tau$-leptons production, the factor $\hat q/m_{\tau}$ remains and
suppresses the cross section.} This is precisely the case of the cross section
for $e^+ e^-$ pair production considered in Ref.~\cite{pzs6.244}.  Another
difference from Eq.~(37) in Ref.~\cite{pzs6.244} is that Ref.~\cite{pzs6.244}
considers the collision in the laboratory frame where the nucleus is at rest and
$\gamma \equiv \gamma_\text{c.m.s.} = (\gamma_\text{lab} / 2)^{1/2}$.

For a proton-proton collision at the LHC with the energy of 13~TeV,
\begin{equation}
  \sigma(pp (\gamma \gamma) \to pp \mu^+ \mu^-) \approx 0.22~\mu\text{b}.
  \label{pp-nocuts-value}
\end{equation}

\section{Cross section of the $\mu^+ \mu^-$ production with experimental cuts}

\label{s:fiducial}

\subsection{Cut on the invariant mass of the $\mu^+ \mu^-$ pair}

The cut on the invariant mass is trivial to apply: only the limits of the
integration over $s$ in~\eqref{pp-nocuts} have to be changed. For $(2 m_\mu)^2
\le \hat s_\text{min} < s < \hat s_\text{max} \le (2 \hat q \gamma)^2$,
\begin{equation}
  \sigma_\text{fid.}^{(\hat s)}(pp (\gamma \gamma) \to pp \mu^+ \mu^-)
  = \int\limits_{\hat s_\text{min}}^{\hat s_\text{max}}
      \mathrm{d} s \,
      \sigma(\gamma \gamma \to \mu^+ \mu^-)
      \int\limits_{s / (2 \hat q \gamma)^2}^{(2 \hat q \gamma)^2 / s}
        \frac{\mathrm{d} x}{8 x} \,
        n \left( \sqrt{\frac{sx}{4}} \right)
        n \left( \sqrt{\frac{s}{4x}} \right).
  \label{pp/s}
\end{equation}
When $\hat s_\text{min} \gg 4 m_\mu^2$, which is valid for the experiments
considered in Section~\ref{s:experiments}, a simplified formula for the
Breit-Wheeler cross section can be used:
\begin{equation}
  \sigma(\gamma \gamma \to \mu^+ \mu^-)
  \approx \frac{4 \pi \alpha^2}{s} \left( \ln \frac{s}{m_\mu^2} - 1 \right)
  \text{ for }
  s \gg 4 m_\mu^2.
\end{equation}
In this case
\begin{equation}
  \sigma_\text{fid.}^{(\hat s)}(pp (\gamma \gamma) \to pp \mu^+ \mu^-)
  = \frac{64 \alpha^4}{3 \pi}
    \int\limits_{\hat s_\text{min}}^{\hat s_\text{max}}
       \frac{\mathrm{d} s}{s^2}
       \left( \ln \frac{s}{m_\mu^2} - 1 \right)
       \ln^3 \frac{2 \hat q \gamma}{\sqrt{s}}.
 \label{pp/s-ll}
\end{equation}
According to Eq.~(6.27b) from~\cite{prep15.181}, the inaccuracy of this formula
originating from virtuality of the photons equals
\begin{equation}
  \eta \sim \left( \frac{\hat q^2}{\sqrt{s_\text{min}} m_\mu} \right)^2
            \left( \ln \frac{4 E^2}{s_\text{min}} \right)^{-1},
\end{equation}
where $E$ is the energy of the colliding particles. The accuracy is very high
for muon-antimuon pair production, but it is considerably worse in the case of
electron-positron pair production.


\subsection{Cut on the muon transverse momentum}

To apply the cut on muon transverse momentum $p_T > \hat p_T$, an expression for
the differential cross section of the $\gamma \gamma \to \mu^+ \mu^-$ reaction
with respect to $p_T$ should be substituted
into~\eqref{pp/s}~\cite[Eq.~(88.4)]{landau-lifshitz-4}:
\begin{equation}
  \mathrm{d} \sigma(\gamma \gamma \to \mu^+ \mu^-)
  = \frac{2 \pi \alpha^2}{s^2}
    \left( \frac{s + t}{t} + \frac{t}{s + t} \right)
    \mathrm{d} t
  = \frac{8 \pi \alpha^2}{s p_T}
    \frac{1 - 2 p_T^2 / s}{\sqrt{1 - 4 p_T^2 / s}}
    \, \mathrm{d} p_T,
\end{equation}
where $t$ is the Mandelstam variable, $t = -s/2 \pm s/2 \cdot \sqrt{1 - 4 p_T^2
/ s}$, and muons are assumed to be ultrarelativistic. The resulting expression
is
\begin{align}
  \sigma_\text{fid.}^{(\hat s, \hat p_T)}(pp (\gamma \gamma) \to pp \mu^+ \mu^-)
  &= \int\limits_{\hat s_\text{min}}^{\hat s_\text{max}}
       \mathrm{d} s
       \int\limits_{\hat p_T}^{\sqrt{s} / 2}
         \mathrm{d} p_T \,
         \frac{\mathrm{d} \sigma(\gamma \gamma \to \mu^+ \mu^-)}{\mathrm{d} p_T}
         \int\limits_{s / (2 \hat q \gamma)^2}^{(2 \hat q \gamma)^2 / s}
           \frac{\mathrm{d} x}{8 x} \,
           n \left( \sqrt{\frac{sx}{4}} \right)
           n \left( \sqrt{\frac{s}{4x}} \right)
  \\
  &= \frac{64 \alpha^4}{3 \pi}
     \int\limits_{\hat s_\text{min}}^{\hat s_\text{max}}
       \frac{\mathrm{d} s}{s^2}
       \ln^3 \frac{2 \hat q \gamma}{\sqrt{s}}
       \left(
         \ln \frac{1 + \sqrt{1 - 4 \hat p_T^2 / s}}
                  {1 - \sqrt{1 - 4 \hat p_T^2 / s}}
         - \sqrt{1 - \frac{4 \hat p_T^2}{s}}
       \right).
   \label{pp/s-pt}
\end{align}

\subsection{Cut on the muon pseudorapidity}

Pseudorapidity is defined as $\eta = -\ln \tan(\theta / 2)$, where $\theta$ is
the angle between the momentum of the muon and the beam axis.  Experimental cuts
on pseudorapidity are related to the detector geometry. The muon spectrometer of
the ATLAS experiment is unable to detect muons with $\theta \lesssim 10^\circ$
or $\theta \gtrsim 170^\circ$, hence the pseudorapidity cut $\abs{\eta} < 2.4$.

For a given value of the muon pair invariant mass $s$, muon pseudorapidities are
determined by the ratio of photon energies $x$. For $x = 1$ and for cuts on
$p_T$ and $s$ implemented in Ref.~\cite{1708.04053} (see Table~\ref{t:pp-cuts}),
$\sin \theta = 2 p_T / \sqrt{s}$ is always larger than $2/7$. Thus $17^\circ
\lesssim \theta \lesssim 163^\circ$, and the cut on $\eta$ does not reduce the
number of detected muon pairs. However, for $x \ll 1$ or $x \gg 1$ muons
propagate in the direction of the proton beam and escape the detector.  Thus, a
cut on pseudorapidity can be naturally transformed into a cut on~$x$:
\begin{equation}
  \abs{\eta} < \hat \eta \Rightarrow 1 / \hat x < x < \hat x,
\end{equation}
where
\begin{equation}
  \hat x
  = \mathrm{e}^{2 \hat \eta}
  \cdot \frac{1 - \sqrt{1 - 4 p_T^2 / s}}
             {1 + \sqrt{1 - 4 p_T^2 / s}}
\end{equation}
(see Appendix~\ref{s:eta->x} for derivation).
In this case the expression for the fiducial cross section is
\begin{align}
  \sigma_\text{fid}^{(\hat s, \hat p_T, \hat \eta)}(pp (\gamma \gamma) \to pp \mu^+ \mu^-)
  &= \int\limits_{\hat s_\text{min}}^{\hat s_\text{max}}
       \mathrm{d} s
       \int\limits_{\hat p_T}^{\sqrt{s} / 2}
         \mathrm{d} p_T
         \frac{\mathrm{d} \sigma(\gamma \gamma \to \mu^+ \mu^-)}{\mathrm{d} p_T}
         \int\limits_{1 / \hat x}^{\hat x}
           \frac{\mathrm{d} x}{8 x}
           n \left( \sqrt{\frac{sx}{4}} \right)
           n \left( \sqrt{\frac{s}{4x}} \right)
  \label{pp-n/s-pt-eta}
  \\
  &= \frac{4 \alpha^4}{\pi}
     \int\limits_{\hat s_\text{min}}^{\hat s_\text{max}}
       \frac{\mathrm{d} s}{s^2}
       \int\limits_{\hat p_T}^{\sqrt{s} / 2}
         \frac{\mathrm{d} p_T}{p_T}
         \frac{1 - 2 p_T^2 / s}{\sqrt{1 - 4 p_T^2 / s}}
         \int\limits_{1 / \hat x}^{\hat x}
           \frac{\mathrm{d} x}{x}
           \ln \frac{(2 \hat q \gamma)^2}{sx}
           \ln \left( \frac{(2 \hat q \gamma)^2}{s} \cdot x \right).
  \label{pp/s-pt-eta}
\end{align}

\section{Comparison with the experimental data}

\label{s:experiments}

\subsection{Muon pair production in proton-proton collisions}

The ATLAS collaboration has measured the fiducial cross section of the $pp \to
pp \mu^+ \mu^-$ reaction at collision energy equal to 13~TeV ($\gamma = 6.93
\cdot 10^3$) with integrated luminosity
$3.2~\text{fb}^{-1}$~\cite{1708.04053}. The experimental cuts are described in
Table~\ref{t:pp-cuts}. The experimental result is
\begin{equation}
  \sigma_\text{fid.}^\text{(exp.)}(pp \to pp \mu^+ \mu^-)
  = 3.12 \pm 0.07~\text{(stat.)} \pm 0.10~\text{(syst.)~pb}.
  \label{pp-experiment}
\end{equation}

\begin{figure}[!tbh]
  \centering
  \includegraphics{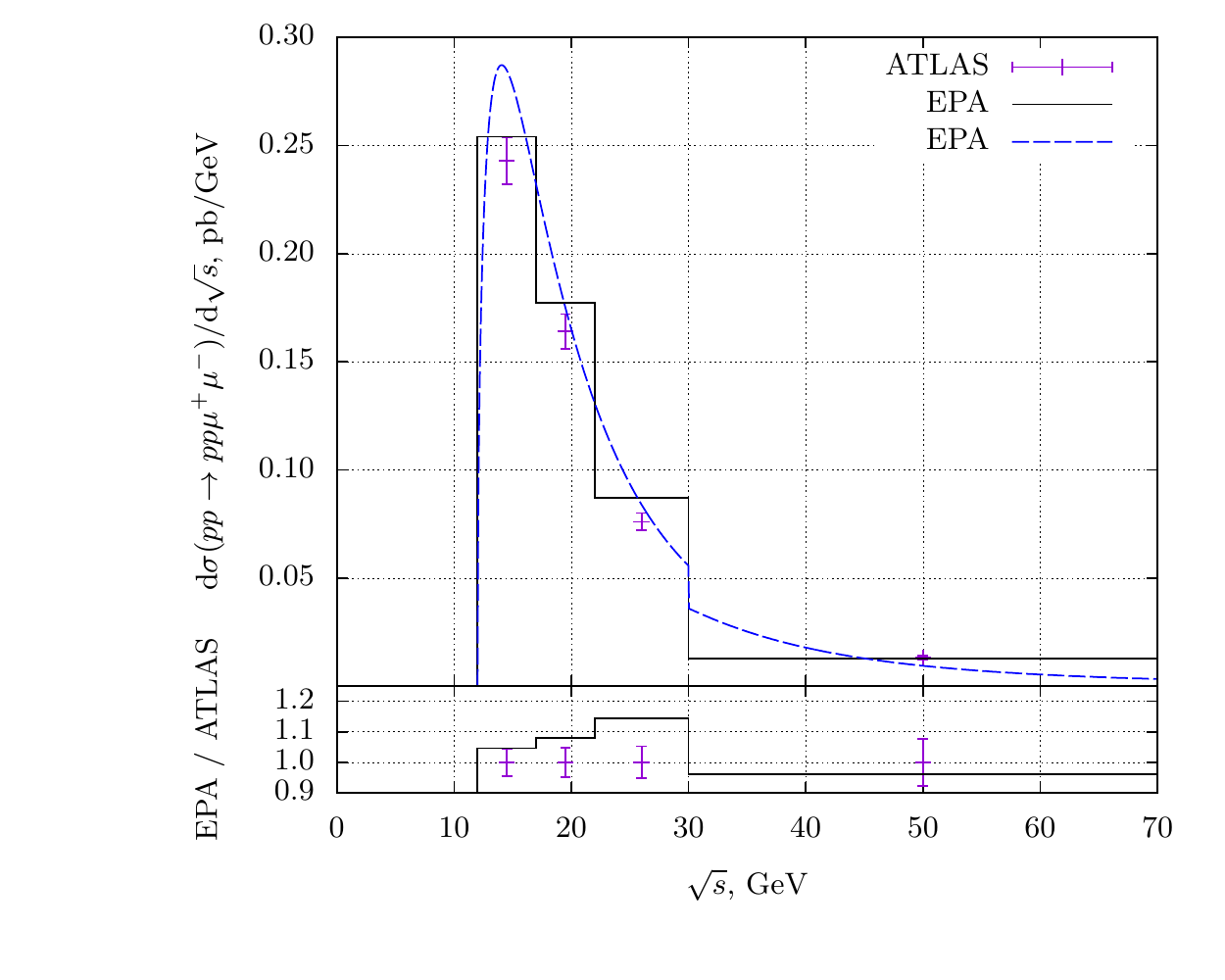}
  \caption{\emph{Upper plot:} fiducial cross section for the $pp (\gamma \gamma)
    \to pp \mu^+ \mu^-$ reaction at proton-proton collision energy 13~TeV
    with the cuts described in Table~\ref{t:pp-cuts}. Points are the
    experimental data presented in Table~3 of Ref.~\cite{1708.04053}. The dashed
    line is the differential cross section calculated with the help
    of~\eqref{pp/s-pt-eta}. The histogram is the differential cross section
    averaged according to the bins also presented in Table~3 of
    Ref.~\cite{1708.04053}. \emph{Lower plot:} ratio of the calculated cross
    section to the experimental points.
  }
  \label{f:pp-results}
\end{figure}

Results of successive application of cuts are presented in
Table~\ref{t:pp-results}. The fiducial cross section is found to be
\begin{equation}
  \sigma_\text{fid.}^{(\hat s, \hat p_T, \hat \eta)}(pp (\gamma \gamma) \to pp \mu^+ \mu^-)
  = 3.35~\text{pb},
\end{equation}
and it is in agreement with the experimental value~\eqref{pp-experiment}.
Fig.~\ref{f:pp-results} compares fiducial cross sections for several bins of
muon pair invariant masses with the experimental data provided in Table~3 of
Ref.~\cite{1708.04053}.\footnote{
  Equivalent photon spectrum~\eqref{spectrum} was used to calculate the
  differential cross section in Fig.~\ref{f:pp-results}. Taking into account
  dipole form factor~\eqref{spectrum-dipole} increases the cross section by less
  than $0.5$\% in the considered energy region. Magnetic form
  factor~\eqref{ff-magnetic} increases the cross section by $\approx 6$\%.
}
The authors of Ref.~\cite{1708.04053} compare their
result with theoretical predictions obtained with the help of Monte Carlo
method: the {\tt SuperCHIC}~\cite{1508.02718} program gives
\begin{equation}
  \sigma_\text{fid.}^{\text{\cite{1708.04053, 1508.02718}}} = 3.45 \pm 0.05~\text{pb};
\end{equation}
EPA prediction corrected for the survival factor~\cite{1410.2983} (see the
discussion in Appendix~\ref{s:survival}) gives
\begin{equation}
  \sigma_\text{fid.}^{\text{\cite{1708.04053, 1410.2983}}} = 3.06 \pm 0.05~\text{pb}.
\end{equation}

\enlargethispage{\baselineskip}

\begin{table}[!tbh]
  \centering
  \caption{Experimental cuts for the fiducial cross section of the $pp (\gamma
  \gamma) \to pp \mu^+ \mu^-$ reaction measured in Ref.~\cite{1708.04053}.}
  \begin{tabular}{ccc}
    Muon pair invariant mass range
    & Muon transverse momentum
    & Muon pseudorapidity
    \\ \hline
    $12~\text{GeV} < \sqrt{s} < 30~\text{GeV}$
    & $p_T > 6$~GeV
    & \multirow{2}{*}{$\abs{\eta} < 2.4$}
    \\
    $30~\text{GeV} < \sqrt{s} < 70~\text{GeV}$
    & $p_T > 10$~GeV
    &
  \end{tabular}
  \label{t:pp-cuts}
\end{table}

\begin{table}[!tbh]
  \centering
  \caption{Fiducial cross sections for the reaction $pp (\gamma \gamma) \to pp
  \mu^+ \mu^-$ calculated with the equivalent photon spectrum~\eqref{spectrum}
  via Eqs.~\eqref{pp-nocuts-ll}, \eqref{pp/s-ll}, \eqref{pp/s-pt} and
  \eqref{pp/s-pt-eta}.}
  \begin{tabular}{lcc}
    \multicolumn{1}{c}{Cuts} & \multicolumn{2}{c}{Cross section, pb}
    \\ \hline \hline
    No cuts & \multicolumn{2}{c}{$1.7 \cdot 10^5$ \rule{0pt}{2.2ex}} \\ \hline
    $12~\text{GeV} < \sqrt{s} < 30~\text{GeV}$
    & $54.1$
    & \multirow{2}{*}{$59.7$}
    \\
    $30~\text{GeV} < \sqrt{s} < 70~\text{GeV}$
    & $5.66$
    &
    \\ \hline
    $12~\text{GeV} < \sqrt{s} < 30~\text{GeV}$,
    $p_T > 6$~GeV
    & $5.38$
    & \multirow{2}{*}{$6.29$}
    \\
    $30~\text{GeV} < \sqrt{s} < 70~\text{GeV}$,
    $p_T > 10$~GeV
    & $0.91$
    &
    \\ \hline
    $12~\text{GeV} < \sqrt{s} < 30~\text{GeV}$,
    $p_T > 6$~GeV,
    $\abs{\eta} < 2.4$
    & $2.85$
    & \multirow{2}{*}{$3.35$}
    \\
    $30~\text{GeV} < \sqrt{s} < 70~\text{GeV}$,
    $p_T > 10$~GeV,
    $\abs{\eta} < 2.4$
    & $0.50$
    &
  \end{tabular}
  \label{t:pp-results}
\end{table}

\clearpage

\subsection{Muon pair production in lead-lead collisions}

\label{s:lead}

The ATLAS collaboration has measured the fiducial cross section of the
$\text{Pb} \; \text{Pb} \to \text{Pb} \; \text{Pb} \; \mu^+ \mu^-$ reaction
at collision energy per nucleon pair equal to $5.02$~TeV ($\gamma = 2.69 \cdot
10^3$) with integrated luminosity
$515~\mu\text{b}^{-1}$~\cite{atlas-conf-2016-025}. The experimental cuts are:
\begin{itemize}
  \item Muon pair invariant mass range: $10~\text{GeV} < \sqrt{s} <
  100~\text{GeV}$.
  \item Muon transverse momentum: $p_T > 4~\text{GeV}$.
  \item Muon pseudorapidity: $\abs{\eta} < 2.4$.
\end{itemize}
The experimental result is
\begin{equation}
  \sigma_\text{fid.}^\text{(exp.)}(\text{Pb} \; \text{Pb} \to \text{Pb} \;
  \text{Pb} \; \mu^+ \mu^-)
  = 32.2 \pm 0.3~\text{(stat.)}^{+4.0}_{-3.4}~\text{(syst.)}~\mu\text{b}.
  \label{pbpb-experiment}
\end{equation}

A heavy nucleus is easier to break apart than a proton. Maximum momentum
transfer for a proton is $\hat q \approx 0.20$~GeV~\eqref{p-qmax}. The
corresponding value for \nuc{Pb}{208} heavily depends on the nucleus form
factor, but it is about an order of magnitude less. In the leading logarithmic
approximation, maximum photon energy is $2 \hat q \gamma$. For the protons with
the collision energy of $13$~TeV, this value is $2.8$~TeV, while for the
lead-lead collision considered in this section, it is about 100~GeV.
Consequently, lead-lead collisions are much more sensitive to the shape of
electromagnetic form factor of colliding particles.

\begin{figure}[!tbh]
  \centering
  \includegraphics{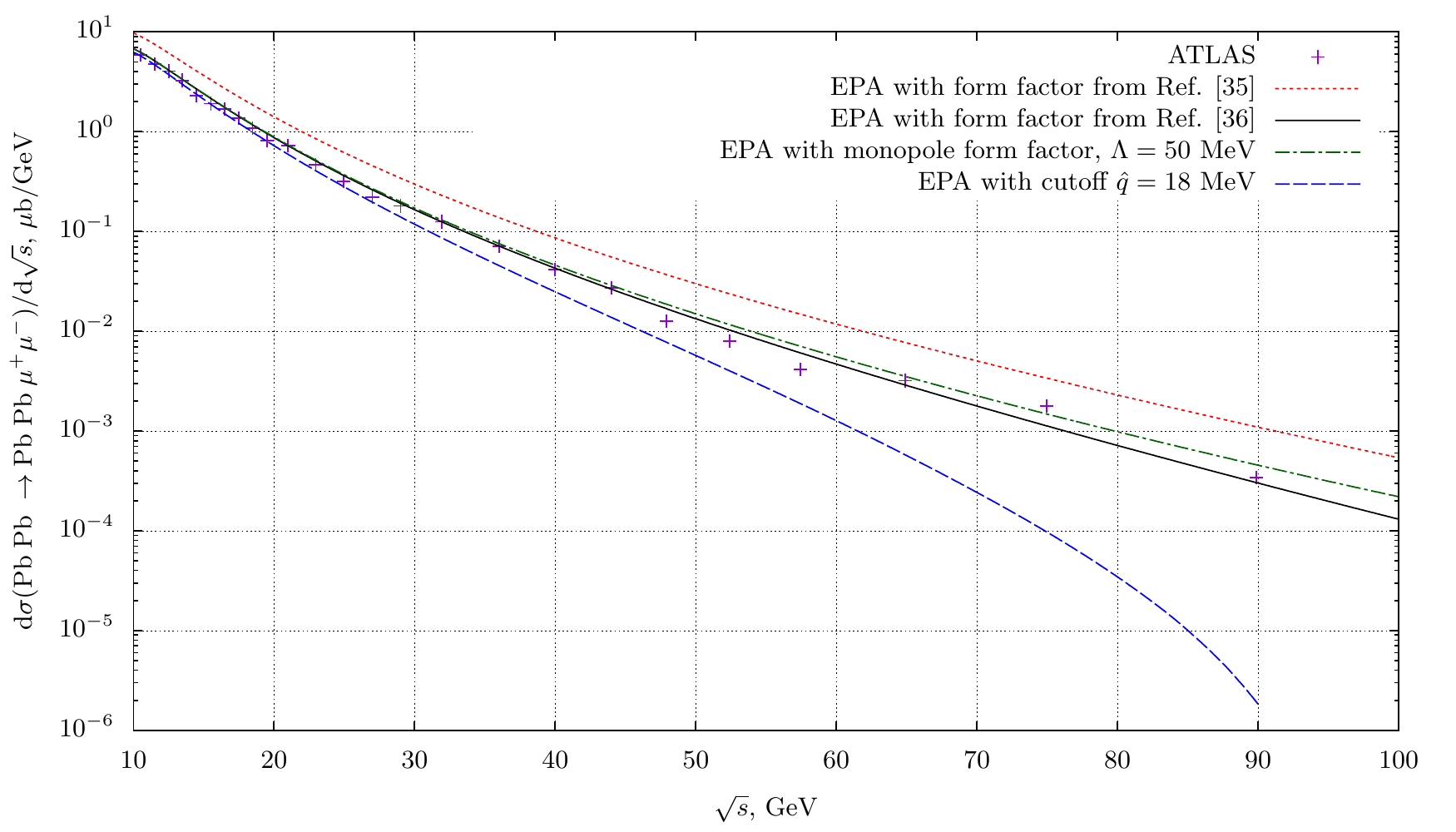}
  \caption{Fiducial cross section for the $\text{Pb} \; \text{Pb} \; (\gamma
  \gamma) \to \text{Pb} \; \text{Pb} \; \mu^+ \mu^-$ reaction at collision
  energy per nucleon pair $5.02$~TeV with the experimental cuts of
  Ref.~\cite{atlas-conf-2016-025} (also described in the text). Points are experimental
  data from the left plot of Fig.~3 of Ref.~\cite{atlas-conf-2016-025} (the
  upper curve). The lines were calculated with the help
  of~\eqref{pp-n/s-pt-eta}.  The red dotted and black solid lines are calculated
  using the equivalent photon spectra with form factors from
  Refs.~\cite{adndt36-495} and~\cite{adndt60-177} correspondingly. The green
  dash-dotted line corresponds to the spectrum with monopole form factor with
  the parameter $\Lambda = 50$~MeV. For the blue dashed line the
  spectrum~\eqref{spectrum} was used with $\hat q = 18$~MeV.}
  \label{f:pbpb-results}
\end{figure}

To calculate the fiducial cross sections, Eq.~\eqref{pp-n/s-pt-eta} was used with
several equivalent photon spectra $n(\omega)$ corresponding to different form
factors~\cite{adndt36-495, adndt60-177}.  Fig.~\ref{f:pbpb-results} compares the
results with the experimental data presented in the left plot in Fig.~3 of
Ref.~\cite{atlas-conf-2016-025}.\footnote{
  The two sets of data points in the left plot in Fig.~3 of
  Ref.~\cite{atlas-conf-2016-025} are for two cuts on dimuon pair rapidity
  $Y_{\mu \mu}$. The cut on dimuon pair rapidity is not considered in this
  paper.  The cut $\abs{Y_{\mu \mu}} < 2.4$ used for the upper curve corresponds
  to the cut $\abs{\eta} < 2.4$.
} The spectrum with the form factor described by Fourier-Bessel parameters from
Ref.~\cite{adndt60-177} (see Table~\ref{t:ff-fourier-bessel}) and the spectrum
with the monopole form factor with the parameter $\Lambda = 50$~MeV both
describe the experimental data well. The leading logarithmic approximation with
$\hat q = 18$~MeV~\eqref{qhat-monopole} is accurate at low invariant masses, but
at high invariant masses it underestimates the number of equivalent photons.
The reason is that in this region the assumption $\omega \ll \hat q \gamma
\approx 50$~GeV used in the derivation of Eq.~\eqref{spectrum} is not valid.
The form factor described by Fourier-Bessel parameters in earlier
publication~\cite{adndt36-495} (see Table.~\ref{t:ff-fourier-bessel}) and its
approximation with the monopole form factor with the parameter $\Lambda =
80$~MeV often used in literature~\cite{prc75-034903, 0901.0891, 0908.3853}
result in the fiducial cross section about $1.5$ times larger than measured.

Fiducial cross section calculated with the spectrum with the form factor
obtained from Fourier-Bessel parameters from Ref.~\cite{adndt60-177},
\begin{equation}
  \sigma_\text{fid.}^{(\hat s, \hat p_T, \hat \eta)}
  (\text{Pb} \; \text{Pb} \; (\gamma \gamma) \to \text{Pb} \; \text{Pb} \; \mu^+ \mu^-)
  = 34.4~\mu\text{b},
\end{equation}
is in agreement with the experimental value~\eqref{pbpb-experiment}.  Cross
sections calculated with successive application of the cuts are summarized in
Table~\ref{t:pbpb-results}.

The authors of Ref.~\cite{atlas-conf-2016-025} compare the measured result with
calculations with the help of the STARLIGHT program~\cite{starlight}:
\begin{equation}
  \sigma^\text{\cite{atlas-conf-2016-025, starlight}}_\text{fid.}
        (\text{Pb} \; \text{Pb} \; (\gamma \gamma) \to \text{Pb} \; \text{Pb} \; \mu^+ \mu^-)
  = 31.64 \pm 0.04~\mu\text{b}.
\end{equation}

\begin{table}[!tbh]
  \centering
  \caption{Fiducial cross sections for the reaction $\text{Pb} \; \text{Pb} \;
  (\gamma \gamma) \to \text{Pb} \; \text{Pb} \; \mu^+ \mu^-$ with the nucleus
  form factor approximated by the monopole formula~\eqref{ff-monopole} with
  $\Lambda = 50$~MeV.}
  \begin{tabular}{lcc}
    Cuts                                        & Cross section, $\mu$b \\\hline
    No cuts                                     & $1.92 \cdot 10^6$ \rule{0pt}{2.2ex} \\
    $10~\text{GeV} < \sqrt{s} < 100~\text{GeV}$ & 264 \\
    also $p_T > 4$~GeV                          & $42.5$ \\
    also $\abs{\eta} < 2.4$                     & $34.6$
  \end{tabular}
  \label{t:pbpb-results}
\end{table}

\section{Conclusions}

The LHC can be used to search for New Physics appearing in photon-photon
collisions.  Photon pair invariant mass can reach $2 \hat q \gamma \approx
2.8$~TeV in $pp$ collisions with the energy equal to 13~TeV and $\approx
100$~GeV in $\text{Pb} \; \text{Pb}$ collisions with the energy per nucleon pair
equal to $5.02$~TeV.

The equivalent photon approximation permits analytical calculation of fiducial
cross section. Leading logarithmic approximation~\eqref{spectrum} provides both
accurate results and relatively simple expressions at invariant masses much less
than $\hat q \gamma$. At higher invariant masses form factors of the colliding
particles have to be taken into account.

Although experimental cuts greatly reduce the production cross section, high
luminosity achieved at the LHC makes it possible to observe $\mu^+ \mu^-$ pair
production in ultraperipheral collisions.

We are grateful to A.~N.~Rozanov, discussion with whom triggered out interest to
the LHC results on $\gamma \gamma$ reactions, to I.~I.~Tsukerman for useful
comments, to H.~Terazawa for drawing our attention to
papers~\cite{rev.mod.phys.4.615, prd4.1532}, to I.~F.~Ginzburg for a very
useful discussion, to S.~I.~Godunov for the help with checking our numerical
calculations, and to V.~A.~Novikov for the ideas implemented in
Appendix~\ref{s:survival}.  We were supported by the RFBR grant 16-02-00342.
Work on Sections \ref{s:xsection}, \ref{s:fiducial} and
Appendix~\ref{s:survival} was supported by the RSF grant 19-12-00123.

\appendix
\numberwithin{equation}{section}

\section{Equivalent photons momentum cutoff}

\label{s:form-factor}

Consider a charged particle at rest. Its electromagnetic field can be
interpreted as a collection of virtual photons with zero energy. Let $q = (0,
q_x, q_y, q_z)$ be the momentum of such a virtual photon. When the particle is
boosted with the Lorentz factor $\gamma \gg 1$ along the $z$ axis, the photon
acquires energy $\omega = \sqrt{\gamma^2 - 1} \, q_z$ approximately equal to the
photon momentum in the boost direction $\gamma q_z$. The virtuality of such a
photon, $-q^2 = q_x^2 + q_y^2 + q_z^2 \ll \omega^2$, so the photon can be
considered real, and this is the essence of the equivalent photon approximation.

To obtain the spectrum of virtual photons $n(\omega)$~\eqref{spectrum} of a
moving particle, the distribution of virtual photons $n(\vec q) \, \mathrm{d}^3
q$~\eqref{epa-distribution} has to be integrated over the photon transverse
momentum $q_\perp = \sqrt{q_x^2 + q_y^2}$. This integral is logarithmically
divergent at high $q_\perp$, and a cutoff is required. In a collision, if a
proton (or a nucleus) emits a virtual photon of sufficiently high momentum, the
proton breaks apart. Thus, a natural estimation for the cutoff value $\hat q$
would be the inverse radius of the proton or the QCD scale $\Lambda_\text{QCD}$
which is in the range of 200--300~MeV~\cite[Section~9]{pdg}.  In the case of
$e^+ e^-$ pair production, $\hat q = m_e$ since contribution of the $q_\perp >
m_e$ domain is power suppressed.

A more rigorous approach to obtain the cutoff value $\hat q$ for proton is to
consider the proton form factor. The~Dirac form factor is~\cite{phys.rep.550.1}
\begin{equation}
  F_1(q^2) = \frac{G_E(q^2) + \tau G_M(q^2)}{1 + \tau},
  \label{form-factor-1}
\end{equation}
where $\tau = -q^2 / 4 m_p^2$,
\begin{equation}
  G_E(q^2) = \frac{1}{(1 - q^2 / \Lambda^2)^2}
\end{equation}
is the electric form factor,
\begin{equation}
  G_M(q^2) = \frac{\mu_p}{(1 - q^2 / \Lambda^2)^2}
  \label{ff-magnetic}
\end{equation}
is the magnetic form factor, $\mu_p = 2.79$ is the proton magnetic moment and
$\Lambda^2 = 0.71~\text{GeV}^2$. Eq.~\eqref{form-factor-1} can be rearranged as
follows
\begin{equation}
  F_1(q^2) = G_D(q^2) \left[ 1 + \frac{(\mu_p - 1) \tau}{1 + \tau} \right],
\end{equation}
where $G_D(q^2) \equiv G_E(q^2)$ is the dipole form factor. Since $-q^2 \approx
q_\perp^2$ cannot be much larger than $\Lambda_\text{QCD}^2$, $\tau \ll 1$ and
the contribution from the magnetic form factor can be neglected. Deriving the
equivalent photon momentum distribution~\eqref{epa-distribution} according
to~\cite[\textsection 99]{landau-lifshitz-4} and taking into account the form
factor results in
\begin{equation}
  n_\text{dipole}(\vec q) \mathrm{d}^3 q
  = \frac{\alpha}{\pi^2}
    \frac{\vec q_{\perp}^{\, 2}}{\omega q^4}
    \left( 1 - \frac{q^2}{\Lambda^2} \right)^{-4}
    \mathrm{d}^3 q
  = \frac{\alpha}{\pi^2 \omega}
    \frac{\vec q_\perp^{\, 2}}{(\omega^2 / \gamma^2 + q_\perp^2)^2}
    \left(
      1
      + \frac{1}{\Lambda^2}
        \left( \frac{\omega^2}{\gamma^2} + q_\perp^2 \right)
    \right)^{-4}
    \mathrm{d}^3 q.
\end{equation}
The equivalent photon spectrum
\begin{equation}
  n_\text{dipole}(\omega) \mathrm{d} \omega
  = 2 \pi
    \int\limits_0^\infty
      n_\text{dipole}(\vec q) q_\perp \mathrm{d} q_\perp \mathrm{d} \omega
  = \frac{\alpha}{\pi}
    \left[
      (4a + 1) \ln \left(1 + \frac{1}{a} \right)
      - \frac{24 a^2 + 42 a + 17}{6 (a + 1)^2}
    \right]
    \frac{\mathrm{d} \omega}{\omega},
  \label{spectrum-dipole}
\end{equation}
where $a = (\omega / \Lambda \gamma)^2$. This function monotonically decreases
with $\omega$. In the lower energy limit $\omega \ll \Lambda \gamma$, where most
of the photons reside,
\begin{equation}
  n_\text{dipole}(\omega) \mathrm{d} \omega
  \xrightarrow[a \to 0]{}
  \frac{\alpha}{\pi}
  \left[ 2 \ln \frac{\Lambda \gamma}{\omega} - \frac{17}{6} \right]
  \frac{\mathrm{d} \omega}{\omega}.
\end{equation}
Comparing this expression with Eq.~\eqref{spectrum} for $Z = 1$, we get
\begin{equation}
  \hat q = \Lambda \mathrm{e}^{-\frac{17}{12}} \approx 0.20~\text{GeV},
  \label{p-qmax}
\end{equation}
which is in a perfect agreement with the previous assumption that $\hat q
\approx \Lambda_\text{QCD}$.


\begin{table}[!b]
  \centering
  \caption{Parameters of the Fourier-Bessel decomposition of \nuc{Pb}{208}
  form-factor~\eqref{ff-fourier-bessel}.}
  \label{t:ff-fourier-bessel}
  \begin{minipage}{\textwidth}
    \centering
    \begin{tabular}{c|r@{$\times$}l|r@{$\times$}l}
      Ref.
      & \multicolumn{2}{c|}{\cite{adndt36-495}\footnote{
          There are two sets of parameters in Ref.~\cite{adndt36-495}. The
          corresponding form factors almost coincide.
        }}
      & \multicolumn{2}{c}{\cite{adndt60-177}} \\ \hline
      $R$, fm
      & \multicolumn{2}{c|}{11.0}
      & \multicolumn{2}{c}{12.5} \\
      $a_1$    & $ 0.62732$ & $10^{-1}$ & $ 1.4396$ & $10^{ 0}$ \\
      $a_2$    & $ 0.38542$ & $10^{-1}$ & $-4.1850$ & $10^{-1}$ \\
      $a_3$    & $-0.55105$ & $10^{-1}$ & $-9.1763$ & $10^{-2}$ \\
      $a_4$    & $-0.26990$ & $10^{-2}$ & $ 6.8006$ & $10^{-2}$ \\
      $a_5$    & $ 0.31016$ & $10^{-1}$ & $ 2.6476$ & $10^{-2}$ \\
      $a_6$    & $-0.99486$ & $10^{-2}$ & $-1.5307$ & $10^{-2}$ \\
      $a_7$    & $-0.93012$ & $10^{-2}$ & $-7.1246$ & $10^{-3}$ \\
      $a_8$    & $ 0.76653$ & $10^{-2}$ & $ 2.7987$ & $10^{-3}$ \\
      $a_9$    & $ 0.20885$ & $10^{-2}$ & $ 2.3767$ & $10^{-3}$ \\
      $a_{10}$ & $-0.17840$ & $10^{-2}$ & $-1.0125$ & $10^{-3}$ \\
      $a_{11}$ & $ 0.74876$ & $10^{-4}$ & $-2.5836$ & $10^{-4}$ \\
      $a_{12}$ & $ 0.32278$ & $10^{-3}$ & $ 6.4297$ & $10^{-5}$ \\
      $a_{13}$ & $-0.11353$ & $10^{-3}$ & $ 6.5528$ & $10^{-5}$ \\
      $a_{14}$ & \multicolumn{2}{c|}{}& $ 1.4523$ & $10^{-5}$ \\
      $a_{15}$ & \multicolumn{2}{c|}{}& $-1.4430$ & $10^{-5}$ \\
    \end{tabular}
 \end{minipage}
\end{table}

\begin{figure}[h]
  \centering
  \includegraphics{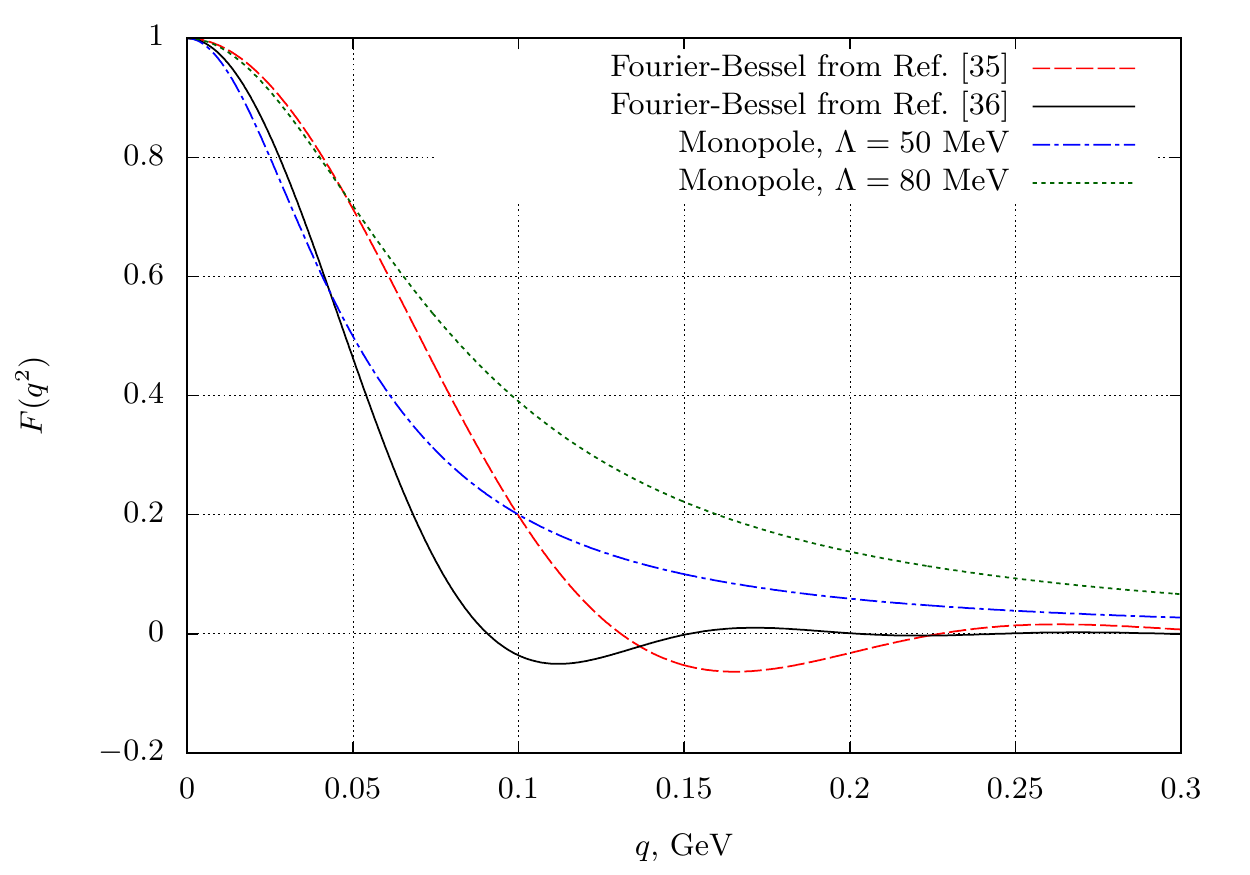}
  \caption{
    \nuc{Pb}{208} form factors available in the literature and their
    approximations. The solid black and the dashed red lines are form factors
    described through Fourier-Bessel decomposition~\eqref{ff-fourier-bessel}.
    Parameters $a_n$ and $R$ for the former were taken from
    Ref.~\cite{adndt60-177}, for the latter---from Ref.~\cite{adndt36-495}. The
    blue dash-dotted and the green dotted lines are monopole
    form-factors~\eqref{ff-monopole} with the parameters $\Lambda = 50$~MeV and
    $\Lambda = 80$~MeV correspondingly.
  }
  \label{f:form-factors}
\end{figure}

Heavy nucleus form factor is more involved. The most accurate description of
the \nuc{Pb}{208} form factor appears to be the Fourier transform of Bessel
decomposition of the nucleus charge density distribution~\cite{npa235-219}:
\begin{equation}
  \rho(r) = \left\{
    \begin{aligned}
      \sum\limits_{n=1}^N a_n j_0 (n \pi r / R) &\text{ for } r \le R, \\
      0 & \text{ for } r \ge R,
    \end{aligned}
  \right.
\end{equation}
where $j_0(x) = \sin(x) / x$ is the Bessel function of order zero, and the
values of $a_n$ and $R$ are provided in Table~\ref{t:ff-fourier-bessel}. The
corresponding form factor is
\begin{equation}
  F_\text{Fourier-Bessel}(q^2)
  = \frac{
     \int \rho(r) \mathrm{e}^{i \vec q \vec r} \mathrm{d}^3 r
    }{
     \int \rho(r) \, \mathrm{d}^3 r
    }
  = \frac{\sin q R}{q R}
  \cdot
    \frac{\sum\limits_{n=1}^N \frac{(-1)^n a_n}{n^2 \pi^2 - q^2 R^2}}
         {\sum\limits_{n=1}^N \frac{(-1)^n a_n}{n^2 \pi^2}}.
  \label{ff-fourier-bessel}
\end{equation}
Heavy nucleus form factor is often approximated by a monopole formula:
\begin{equation}
  F_\text{monopole}(q^2) \approx \frac{1}{1 - q^2 / \Lambda^2}.
  \label{ff-monopole}
\end{equation}
The corresponding equivalent photon spectrum is
\begin{equation}
  n_\text{monopole}(\omega) \mathrm{d} \omega =
  \frac{Z^2 \alpha}{\pi}
  \left[ (2 a + 1) \ln \left( 1 + \frac{1}{a} \right) - 2 \right]
  \frac{\mathrm{d} \omega}{\omega}.
\end{equation}
In the low energy limit
\begin{equation}
  n_\text{monopole}(\omega) \mathrm{d} \omega
  \xrightarrow[a \to 0]{}
  \frac{Z^2 \alpha}{\pi}
  \left[ 2 \ln \frac{\Lambda \gamma}{\omega} - 2 \right]
  \frac{\mathrm{d} \omega}{\omega},
\end{equation}
so
\begin{equation}
  \hat q = \Lambda \mathrm{e}^{-1}.
  \label{qhat-monopole}
\end{equation}
For $\Lambda = 80$~MeV that was used in~\cite{prc75-034903, 0901.0891,
0908.3853}, $\hat q \approx 30$~MeV. However, this value of $\Lambda$ apparently
approximates outdated data. Fig.~\ref{f:form-factors} compares monopole form
factor with $\Lambda = 80$~MeV to form factors calculated through Fourier-Bessel
decomposition with the parameters that were fit to the experimental data
available in~1987~\cite{adndt36-495} and in~1995~\cite{adndt60-177} (see
Table~\ref{t:ff-fourier-bessel}). Monopole form factor with $\Lambda = 50$~MeV
($\hat q \approx 18$~MeV) used in Section~\ref{s:lead} is presented as well for
the reference.

\section{Pseudorapidity cut}

\label{s:eta->x}

\begin{figure}[!tbh]
  \centering


  \begin{tikzpicture}[scale=2]
    \draw [midarrow]
      (0,  1) node [above] {$\gamma$} -- node [left] {$\omega_1$}   (0, 0);
    \draw [midarrow]
      (0, -1) node [below] {$\gamma$} -- node [right] {$\omega_2$}  (0, 0);
    \draw [->] 
      (0,  0) -- node [above left] {$p^-$} (-0.8,-0.6) node [below left] {$\mu^-$};
    \draw [->]
      (0,  0) -- node [below right] {$p^+$} (0.8,1) node [above right] {$\mu^+$};
    \draw
      (0,  0.25) arc (90:51.3:0.25)
      node at (60:0.25) [yshift=3.0ex, xshift=1.0ex] {$\theta_1$};
    \draw
      (0, -0.25) arc (-90:-143.1:0.25)
      node at (-110:0.25) [yshift=-2ex, xshift=-1.5ex] {$\theta_2$};
  \end{tikzpicture}

  \caption{$\gamma \gamma \to \mu^+ \mu^-$ reaction.}
  \label{f:γγ-collision}
\end{figure}
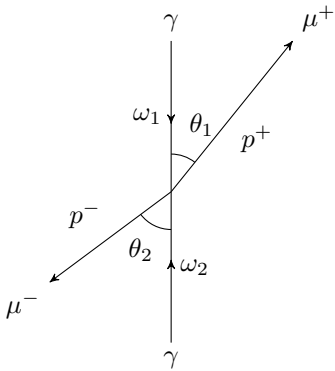

In order to take the pseudorapidity cut into account, the photon energy ratio $x
= \omega_1 / \omega_2$ has to be expressed through the muon pair invariant mass
$s$, muon transverse momentum $p_T$, and muon pseudorapidity $\eta$. A collision
of two photons with the energies $\omega_1$ and $\omega_2$ is shown in
Fig.~\ref{f:γγ-collision}.  $\mu^+$ with momentum $p^+$ and $\mu^-$ with
momentum $p^-$ are produced in this collision. In the following $p_T \gg m_\mu$
is assumed, and the muon mass $m_\mu$ is neglected; this is valid for the
experiments considered in this paper.

From the conservation of energy and momenta
\begin{equation}
  \left\{
    \begin{aligned}
      p_T^+ &= -p_T^- \equiv p_T, \\
      \omega_1 + \omega_2
      &= \sqrt{p_T^2 + p_\parallel^{+2}} + \sqrt{p_T^2 + p_\parallel^{-2}},
      \\
      \omega_1 - \omega_2 &= p_\parallel^- - p_\parallel^+.
    \end{aligned}
  \right.
\end{equation}
The last two equations can be expressed through the transverse momentum $p_T$
and the scattering angles $\theta_1$ and $\theta_2$:
\begin{equation}
  \left\{
    \begin{aligned}
      \frac{p_T}{\sin \theta_1} + \frac{p_T}{\sin \theta_2}
      &= \omega_1 + \omega_2, \\
      \frac{p_T}{\tan \theta_1} - \frac{p_T}{\tan \theta_2}
      &= \omega_1 - \omega_2.
    \end{aligned}
  \right.
\end{equation}
The scattering angles are related to pseudorapidity through equation
\begin{equation}
  \eta_i = -\ln \tan (\theta_i / 2), \ i = 1,~2,
\end{equation}
so
\begin{equation}
  \left\{
    \begin{aligned}
      \cosh \eta_1 + \cosh \eta_2 &= \frac{\omega_1 + \omega_2}{p_T}, \\
      \sinh \eta_1 - \sinh \eta_2 &= \frac{\omega_1 - \omega_2}{p_T}.
    \end{aligned}
  \right.
\end{equation}
Elimination of $\eta_2$ results in the equation
\begin{equation}
  \mathrm{e}^{2 \eta_1}
  - \frac{2 \omega_1}{p_T} \mathrm{e}^{\eta_1}
  + \frac{\omega_1}{\omega_2}
  = 0.
\end{equation}
Substitution of $\omega_1 = \sqrt{sx/4}$, $\omega_2 = \sqrt{s/4x}$ leads to
\begin{equation}
  \mathrm{e}^{2 \eta_1}
  - \frac{\sqrt{sx}}{p_T} \mathrm{e}^{\eta_1}
  + x
  = 0.
\end{equation}
The solution of this equation with respect to $x$ is
\begin{equation}
  x
  = \mathrm{e}^{2 \eta_1}
  \cdot \frac{(1 \pm \sqrt{1 - 4 p_T^2 / s})^2}{4 p_T^2 / s}
  = \mathrm{e}^{2 \eta_1}
  \cdot \frac{1 \pm \sqrt{1 - 4 p_T^2 / s}}{1 \mp \sqrt{1 - 4 p_T^2 / s}}.
\end{equation}
With $\eta_1$ varying from $-\hat \eta$ to $\hat \eta$, $x$ varies in the
following intervals:
\begin{equation}
  \left\{
    \begin{aligned}
        \mathrm{e}^{-2 \hat \eta}
        \cdot
        \frac{1 + \sqrt{1 - 4 p_T^2 / s}}{1 - \sqrt{1 - 4 p_T^2 / s}}
      < & \, x
      < \mathrm{e}^{2 \hat \eta}
        \cdot
        \frac{1 + \sqrt{1 - 4 p_T^2 / s}}{1 - \sqrt{1 - 4 p_T^2 / s}},
      \\
        \mathrm{e}^{-2 \hat \eta}
        \cdot
        \frac{1 - \sqrt{1 - 4 p_T^2 / s}}{1 + \sqrt{1 - 4 p_T^2 / s}}
      < & \, x
      < \mathrm{e}^{2 \hat \eta}
        \cdot
        \frac{1 - \sqrt{1 - 4 p_T^2 / s}}{1 + \sqrt{1 - 4 p_T^2 / s}}.
    \end{aligned}
  \right.
\end{equation}
To satisfy both $\eta_1 < \abs{\hat \eta}$ and $\eta_2 < \abs{\hat \eta}$, the
intersection of these intervals has to be selected. Hence
\begin{equation}
  1 / \hat x < x < \hat x
  \text{ where }
  \hat x
  = \mathrm{e}^{2 \hat \eta}
  \cdot \frac{1 - \sqrt{1 - 4 p_T^2 / s}}{1 + \sqrt{1 - 4 p_T^2 / s}}.
\end{equation}
When applying these inequalities to setup the integration domain for the
equivalent photon approximation, a check that the photon energy does not exceed
the cutoff energy $\hat q \gamma$ is required:
\begin{equation}
  \hat x < \frac{(2 \hat q \gamma)^2}{s}.
  \label{x-cut}
\end{equation}
This is always true for the reaction $pp(\gamma \gamma) \to pp \mu^+ \mu^-$ with
the cuts implemented in~\cite{1708.04053}. However, in the case of $\text{Pb} \;
\text{Pb} \; (\gamma \gamma) \to \text{Pb} \; \text{Pb} \; \mu^+ \mu^-$
reaction with the cuts used in~\cite{atlas-conf-2016-025}, this inequality
provides an additional cut on $x$ which should be accounted for when calculating
the fiducial cross section with cutoff~$\hat q$.

\section{Survival factor}

\label{s:survival}

It is well known that the distribution of equivalent photons given
by~\eqref{spectrum} can be obtained from the classical consideration of the
electromagnetic field of an ultrarelativistic charged particle. The solution of
Maxwell equations for the electromagnetic field induced by an ultrarelativistic
charged particle moving in the direction of $z$ axis
is~\cite[Eq.~(33.2.3)]{akhiezer}
\begin{equation}
  \begin{aligned}
    \vec E_\perp (\vec r, t)
    &= -\frac{i Z e \gamma}{(2 \pi)^3}
       \int
         \frac{\vec q_\perp}{\vec q^{\; 2}}
         \mathrm{e}^{i \vec q_\perp \vec r + i \gamma q_z (z - v t)}
       \mathrm{d}^3 q, 
    \\
    \vec H(\vec r, t) &= \vec v \times \vec E(\vec r, t),
  \end{aligned}
  \label{maxwell}
\end{equation}
where $\vec E(\vec r, t)$ and $\vec H(\vec r, t)$ are electric and magnetic
fields at point $\vec r$ at the moment $t$, $\vec E_\perp$ is the component of
$\vec E$ transverse to the $z$ axis, $v$ is the particle velocity, and $\vec q$
is the Fourier transformation parameter.

Electric field in the $z$ direction is not enhanced by the Lorentz factor
$\gamma$. Thus for $\gamma \gg 1$, $E_\perp \gg \abs{E_z}$ and
the electric field is practically transversal, just as it should be in the case
of real photons. In this way equation~\eqref{maxwell} gives expansion of
electromagnetic fields in terms of monochromatic plane waves moving in the $z$
direction and having frequencies $\omega = v \gamma q_z \approx \gamma
q_z$.

Total flux of the electromangetic energy flowing in the $z$ direction is given
by the Poynting vector component:
\begin{equation}
  \Pi_z
  = \int \mathrm{d}^2 b
      \int\limits_{-\infty}^\infty
        \mathrm{d} t \;
        [ \vec E \times \vec H ]_z,
  \label{poynting}
\end{equation}
where $\vec b \equiv \vec r_\perp$ is the impact parameter for point $\vec r$.
$\Pi_z$ is equal to the total energy of equivalent photons:
\begin{equation}
  \Pi_z = \int\limits_0^\infty \omega \, n(\omega) \, \mathrm{d} \omega.
  \label{poynting-spectrum}
\end{equation}
Substitution of expansion~\eqref{maxwell} into Eq.~\eqref{poynting} results in:
\begin{multline}
  \Pi_z
  = \int \mathrm{d}^2 b
      \int \mathrm{d} t \;
        (E_x^2 + E_y^2)
  = \frac{Z^2 e^2 \gamma^2}{(2 \pi)^6}
    \int \mathrm{d}^2 b \, \mathrm{d} t \, \mathrm{d}^3 q \, \mathrm{d}^3 q'
      \frac{-q_x q'_x - q_y q'_y}{\vec q^{\; 2} \pvec q'^{\; 2}}
  \\  \times
      \exp[ i x (q_x + q'_x) + i y (q_y + q'_y) + i \gamma (z - v t) (q_z + q'_z)]
      \, F(\vec q^{\; 2})
      \, F(\pvec q'^2),
  \label{poynting-expanded}
\end{multline}
where the form factor of the charged particle is taken into account.
Integration of~\eqref{poynting-expanded} over $\mathrm{d}^2 b$ and $\mathrm{d}
t$ results in three delta functions which then remove the integral over
$\mathrm{d}^3 q'$. The final expression for the $z$ component of the Poynting
vector is
\begin{equation}
  \Pi_z
  = \frac{Z^2 \alpha \gamma}{2 \pi^2}
    \int
      \frac{
        \vec q_\perp^{\; 2} \, F^2(\vec q_\perp^{\; 2} + \omega^2 / \gamma^2)
      }{
        (\vec q_\perp^{\; 2} + \omega^2 / \gamma^2)^2
      }
      \, \mathrm{d}^3 q.
  \label{poynting-integrated}
\end{equation}
Changing the integration variable $q_z$ to $\omega$ and comparing the result
with~\eqref{poynting-spectrum}, we obtain:
\begin{equation}
  n(\omega) \, \mathrm{d} \omega
  = \frac{Z^2 \alpha}{\pi^2}
    \int
      \frac{
        \vec q_\perp^{\; 2} \, F^2(\vec q_\perp^{\; 2} + \omega^2 / \gamma^2)
      }{
        (\vec q_\perp^{\; 2} + \omega^2 / \gamma^2)^2
      }
    \, \mathrm{d}^2 q_\perp
    \frac{\mathrm{d} \omega}{\omega},
  \label{spectrum-ff}
\end{equation}
where the extra factor of 2 comes from the fact that when $q_z$ varies from
$-\infty$ to $\infty$, $\omega$ covers the region $[0; \infty)$ twice. The
difference between this expression and~\eqref{spectrum} is that here the form
factor is taken into account.

In order to introduce the notion of survival factor, the integration over
$\mathrm{d}^2 b$ in~\eqref{poynting-expanded} should be postponed until the end.
Let us define $n(b, \omega)$ through the following equation:
\begin{equation}
  n(\omega) = \int n(b, \omega) \, \mathrm{d}^2 b.
  \label{survival-spectrum-definition}
\end{equation}
Integrating~\eqref{poynting-expanded} over $\mathrm{d} t$ and performing
integration over $\mathrm{d} q'_z$ with the help of the delta function, we
obtain:
\begin{equation}
  n(b, \omega)
  = \frac{Z^2 \alpha}{4 \pi^4 \omega}
    \int \mathrm{d} q_\perp \mathrm{d} \theta \,
    \frac{q_\perp F(q_\perp^2 + \omega^2 / \gamma^2)}
         {q_\perp^2 + \omega^2 / \gamma^2}
    \int \mathrm{d} q'_\perp \mathrm{d} \theta' \,
    \frac{q'_\perp F(q'^2_\perp + \omega^2 / \gamma^2)}
         {q'^2_\perp + \omega^2 / \gamma^2}
    (- q_\perp q'_\perp)
    \cos(\theta - \theta')
    \, \mathrm{e}^{i b q_\perp \cos \theta}
    \, \mathrm{e}^{i b q'_\perp \cos \theta'},
\end{equation}
where $\theta$ is the angle between $\vec q_\perp$ and $\vec b$, and $\theta'$
is the angle between $\pvec q'_\perp$ and $\vec b$.

The next step is to use the integral representation of the Bessel function:
\begin{equation}
  \begin{aligned}
    \int\limits_0^{2 \pi}
    \cos \theta \, \mathrm{e}^{i a \cos \theta} \mathrm{d} \theta
    &= 2 i \pi J_1(a),
    \\
    \int\limits_0^{2 \pi}
    \sin \theta \, \mathrm{e}^{i a \cos \theta} \mathrm{d} \theta
    &= 0,
  \end{aligned}
\end{equation}
and the expansion $\cos(\theta - \theta') = \cos \theta \cos \theta' + \sin
\theta \sin \theta'$:
\begin{equation}
  n(b, \omega)
  = \frac{Z^2 \alpha}{\pi^2 \omega}
     \left[
       \int \mathrm{d} q_\perp
         q_\perp^2
         \frac{F(q_\perp^2 + \omega^2 / \gamma^2)}
              {q_\perp^2 + \omega^2 / \gamma^2}
        J_1(b q_\perp)
    \right]^2.
 \label{survival-spectrum}
\end{equation}
With the help of the identity
\begin{equation}
  \int\limits_0^{\infty} J_1(a x) J_1(b x) \, x \, \mathrm{d} x
  = \frac{1}{a} \, \delta(a - b),
\end{equation}
transition from~\eqref{survival-spectrum}
through~\eqref{survival-spectrum-definition} to~\eqref{spectrum-ff} is
straightforward.

For a standalone ultrarelativistic charged particle, its finite transversal size
is taken into account by the form factor $F$ which tends to 0 for high
transverse photon momentum $q_\perp$. However, when an ultraperipheral collision
of two ultrarelativistic charged particles is considered, one should factor in
the probability that this collision is indeed ultraperipheral, i.e., that the
particles remain intact after the collision. If the particles are modeled as
black disks, then the neccessary requirement is that their impact parameter $b =
\abs{\vec b_2 - \vec b_1}$ is greater than the sum of their radii. More
elaborate model of the particles interaction can be described through the
function $P(b)$ which is the probability for the particles to remain intact
after passing each other at distance $b$. Then the cross section for production
of a system $X$ in an ultraperipheral collision is
\begin{equation}
  \sigma(N N \to N N X)
  = \int\limits_0^\infty \mathrm{d} \omega_1
    \int\limits_0^\infty \mathrm{d} \omega_2
    \, \sigma(\gamma \gamma \to X)
    \int \mathrm{d}^2 b_1
    \int \mathrm{d}^2 b_2
    \, n(b_1, \omega_1)
    \, n(b_2, \omega_2)
    \, P(\abs{\vec b_2 - \vec b_1}),
\end{equation}
where $N$ is the colliding particle, and $\sigma(\gamma \gamma \to X)$ is the
cross section for the production of $X$ through the photon fusion. For
point-like particles $P(b) = 1$, and
\begin{equation}
  \sigma(N N \to N N X)
  = \int\limits_0^\infty \mathrm{d} \omega_1
    \int\limits_0^\infty \mathrm{d} \omega_2
    \, \sigma(\gamma \gamma \to X)
    \, n(\omega_1)
    \, n(\omega_2).
  \label{epa-production}
\end{equation}
(The difference between~\eqref{epa-production} and~\eqref{pp-nocuts-omega} is
that in~\eqref{epa-production} the integration is cut off at high $\omega_i$
through the form factors, while in~\eqref{pp-nocuts-omega} it is done
explicitely through the $\hat q$ parameter.)
In~\cite{1410.2983}, survival factor is defined as 
\begin{equation}
  S_{\gamma \gamma}^2
  = \frac{
      \int\limits_{b_1 > R}
      \, \int\limits_{b_2 > R}
      \, n(b_1, \omega_1)
      \, n(b_2, \omega_2)
      \, P(\abs{\vec b_2 - \vec b_1})
      \, \mathrm{d}^2 b_1
      \, \mathrm{d}^2 b_2
    }{
      \int\limits_{b_1 > 0}
      \, \int\limits_{b_2 > 0}
      \, n(b_1, \omega_1)
      \, n(b_2, \omega_2)
      \, \mathrm{d}^2 b_1
      \, \mathrm{d}^2 b_2
    },
\end{equation}
where $R$ is the radius of the colliding particle. The form factor of the
particle $F(\vec q_\perp^2 + \omega^2 / \gamma^2)$ cuts off the
integration in Eq.~\eqref{survival-spectrum} at large $q_\perp$ or,
equivalently, at small $b$. Consequently the integration in the nominator should
not explicitely cut off the regions $b_1, b_2 < R$. Thus the formula for the
survival factor we suggest is
\begin{equation}
  S_{\gamma \gamma}^2
  = \frac{
      \int\limits_{b_1 > 0}
      \, \int\limits_{b_2 > 0}
      \, n(b_1, \omega_1)
      \, n(b_2, \omega_2)
      \, P(\abs{\vec b_2 - \vec b_1})
      \, \mathrm{d}^2 b_1
      \, \mathrm{d}^2 b_2
    }{
      n(\omega_1) \, n(\omega_2)
    }.
\end{equation}

\newcommand{\arxiv}[1]{\href{http://arxiv.org/abs/#1}{arXiv:\nolinebreak[3]#1}}

\end{document}